\documentclass[%
 reprint,
 amsmath,amssymb,
 aps,
]{revtex4-2}

\usepackage{graphicx}
\usepackage{dcolumn}
\usepackage{bm}

\begin{document}

\title{Magnetic Field Tolerant Superconducting Spiral Resonators for Circuit QED}

\author{Mihirangi Medahinne}
\thanks{These authors contributed equally to this work.}
\affiliation{University of Rochester, Rochester, NY, 14627 USA}

\author{Yadav P. Kandel}
\thanks{These authors contributed equally to this work.}
\thanks{Present address: IBM T.J. Watson Research Center, Yorktown Heights, New York 10598, USA}
\affiliation{University of Rochester, Rochester, NY, 14627 USA}

\author{Suraj Thapa Magar}
\affiliation{University of Rochester, Rochester, NY, 14627 USA}

\author{Elizabeth Champion}
\affiliation{University of Rochester, Rochester, NY, 14627 USA}

\author{John M. Nichol}
\affiliation{University of Rochester, Rochester, NY, 14627 USA}

\author{Machiel S. Blok}
\email{machielblok@rochester.edu}
\affiliation{University of Rochester, Rochester, NY, 14627 USA}

\begin{abstract}
We present spiral resonators of thin film niobium (Nb) that exhibit large geometric inductance, high critical magnetic fields and high single photon quality factors. These low loss geometric inductors can be a compelling alternative to kinetic inductors to create high-impedance superconducting devices for applications that require magnetic fields. By varying the spiral pitch, we realize resonators with characteristic impedances ranging from 3.25-7.09 k$\Omega$. We measure the temperature and magnetic field dependent losses and find that the high-impedance resonators maintain an intrinsic quality factor above $\sim 10^5$ for parallel magnetic fields of up to 1 T. These properties make spiral Nb resonators a promising candidate for quantum devices that require circuit elements with high impedance and magnetic field resilience.
\end{abstract}
\date{\today}

\maketitle
\section{Introduction}

High-impedance quantum circuit elements consisting of large inductors with small self-capacitance are an important building block for noise-protected superconducting qubits, quantum sensors, and hybrid quantum systems. For example, large-impedance inductors are used to engineer the inductive energy independent of the charging and Josephson energies leading to delocalized qubit wavefunctions in phase that are more resilient to flux noise \cite{Vladmir2009_Fluxonium, hassani2023_inductivelySchuntedTransmon}. Furthermore, a resonator with high characteristic impedance has large zero-point voltage fluctuations, leading to enhanced coupling between microwave photons and phonons (either using piezoelectric materials \cite{Arrangoiz-Arriola2018_Ctystal_CouplingQuantumCircuitToPiezoelectric} or static electric fields \cite{bozkurt_quantum_2023}), polar molecules \cite{andre_coherent_2006} and single electrons \cite{Wallraff2017_CavityQEDDoubleQD}. 

One common approach to making high-impedance inductors is to enhance the kinetic inductance with disordered superconductors like Nb-N \cite{Niepce2018_HighKineticInductanceNbNnanowaire}, Nb-Ti-N \cite{Kouwenhoven2019_BResilientCPW,Vandersypen2016_KineticInductanceNWs} and Ti-N \cite{shearrow2018_TiN}, granular aluminum \cite{grunhaupt2019_granularAlHighKineticInductance}, or by forming a large array of Josephson junctions \cite{Masluk2012_JJarray}. However, integration of new materials into devices can be challenging because they have to be compatible with existing fabrication processes and experimental operating conditions such as magnetic fields for coupling to spin qubits. As an alternative, a large geometric inductance can be achieved through the mutual inductance between the concentric loops of a planar spiral to realize an impedance that surpasses the superconducting resistance quantum of $h/(2e)^2=6.5 \, k \Omega $ \cite{peruzzo2020_GemometricSI}. 
The impedance of these spiral inductors is determined by their geometry, leading to a high degree of design control and flexibility. Until now, high-impedance spiral resonators were made from superconducting aluminum whose critical field is about 10 mT \cite{caplan_1965_criticalfieldAl,harris_1968_criticalFieldAl}. This is prohibitively small for most quantum information protocols that require magnetic fields \cite{Kouwenhoven2019_BResilientCPW,veldhorst_2014_addressable}. 

Here, we fabricate and characterize high-impedance spiral resonators with Nb thin films which offer large critical temperature and magnetic field resilience. We experimentally extract the ratio of kinetic and geometric inductance to verify that the primary contribution to the resonator inductance is geometric. We optimize our fabrication recipe to realize a high quality factor $Q_{\rm int}$  of $ \sim 10^5$ that is maintained for in-plane magnetic fields up to $\sim$ 1 T. We analyze the contributions of the noise by performing measurements at varying photon numbers and temperature. We find that at low temperatures and low photon numbers, $Q_{\rm int}$ is dominated by Two-Level System (TLS) losses. With increasing photon number, the TLSs saturate and the losses are dominated by other, power-independent losses. 
By exploring additional cleaning steps and better etch processes in spiral fabrication we can potentially mitigate these TLS losses \cite{mcrae_2020_materialsLossMeasurementsMicrowaveResonators}. We expect that $Q_{\rm int}$ can be further improved by performing in-depth analyses of power independent losses and implementing techniques to reduce them during measurement and fabrication \cite{mcrae_2020_materialsLossMeasurementsMicrowaveResonators}.

\section{Design and Fabrication}

\begin{figure*}[ht]
    \centering
   \includegraphics[scale=0.95] {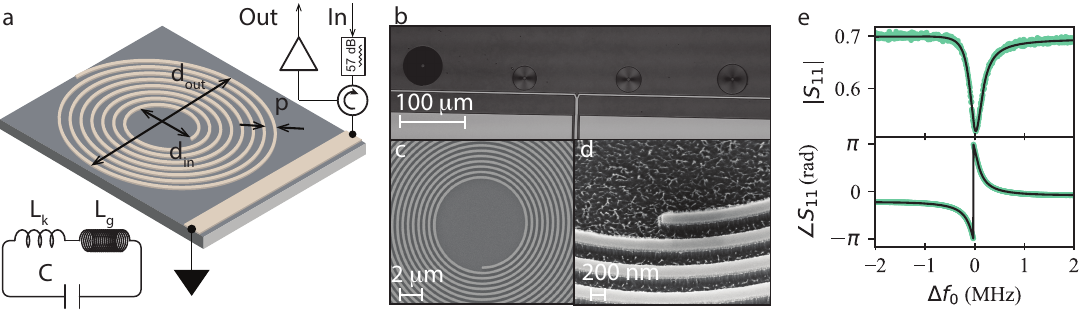}
    \caption{ Experimental setup of the device. (a)  Schematic diagram of the measurement device with a spiral inductively coupled to the CPW feed line. Around its fundamental resonance frequency, a spiral resonator can be modeled as an LC circuit where the total inductance L has contributions from the geometric inductance $L_g$ and the kinetic inductance $L_k$. (b) An optical microscope image of the measurement device (in grayscale). (c-d) Scanning electron microscope (SEM) images of one of the spiral resonators on the measurement device. (e) Magnitude (top) and phase (bottom) of a measured spiral resonator (R0 in the text) at 60 mK. Measured data are shown in green. The magnitude of the measurement has a Lorentzian-shaped dip, and phase has a 2$\pi$ shift. We extract the resonator parameters by fitting a theoretical model for $S_{11}$, shown in black. 
}
    \label{fig:exp_setup}
\end{figure*}
In this work, we design Nb spiral resonators where characteristic impedance $Z_C$ is varied by adjusting the mutual inductance and capacitance between the concentric loops.  At the fundamental resonance frequency $f_{g}$, a planar spiral can be modeled as an  \textit{LC} oscillator, $Z_C =\sqrt{L/C} =2\pi f_{0}L$, where the total inductance $L$ has both kinetic and geometric contributions and $C$ is the capacitance of the spiral [Figure \ref{fig:exp_setup}(a)].  Therefore, by increasing the mutual inductance between the concentric loops of the spiral, we can enhance $Z_{C}$.

An analytic expression for the geometric contribution to the spiral inductance $L_g$, derived with the current sheet method is
 \cite{Mohal_1999_planar_spital_inductances,peruzzo2020_GemometricSI}
\begin{equation}
   L_{g}= \tfrac{\mu_{0}n^{2}d_{\rm av}}{2}\left(ln\left ( 2.5/\rho \right ) + 0.2\rho ^2 \right),
    \label{eq:Lg} 
\end{equation}
where $\mu_{0}$ is the vacuum permeability, $n$ is the number of turns, $d_{\rm av}$ is the average of the inner diameter $d_{\rm in}$ and the outer diameter $d_{\rm out}$, and $\rho$ is the fill ratio of the spiral and it is given as $ \rho= \left ( d_{\rm out}-d_{\rm in} \right )/\left (  d_{\rm out}+d_{\rm in}\right )$ [Figure \ref{fig:exp_setup}(a)]. As $d_{\rm out}$ scales linearly with the number of turns $n$, it can be seen from  Equation \ref{eq:Lg} that $L_g$ scales with the number of turns \textit{n} as $L_{g} \sim n^3 $.
For planar spirals, the analytical formula for the fundamental resonance frequency $f_g$ is \cite{Maleeva_2015_electrodynamics_Archimedean_spiral,peruzzo2020_GemometricSI}
\begin{equation}
 f_{g}= \xi \tfrac{c_0}{\sqrt{\epsilon _{eff}}}\tfrac{2p}{\pi\left ( d_{\rm in} +2np\right )^2},
    \label{eq:fg} 
\end{equation}
where $\xi$ is a shape-dependent constant with $\xi =0.81$  for planar spirals, $c_0$ is the speed of light in vacuum, ${\epsilon _{\rm eff}}$ is the effective dielectric constant, and the pitch $p$ is the wire width plus the gap between adjacent turns of the spiral. Here we can see that $f_g$ scales with the pitch \textit{p} and number of turns $n$ as $f_{0} \sim n^{-2}p^{-1} $ and the impedance $Z_C =2\pi f_{0}L$ scales as  $Z_C \sim p^{-1}n$ \cite{peruzzo2020_GemometricSI}. Because there is no simple geometric parameter that can tune the impedance independently of frequency, we chose to design high-impedance resonators by decreasing the spiral pitch, while adjusting the number of turns to maintain resonance frequencies $f_g$ in the 4-8 GHz range. Smallest pitch used in this work is 300 nm and is limited by our nano-fabrication tools.

We design the spirals with a pitch ranging from 300 nm to 1000 nm with a characteristic impedance between 3-7 k$\Omega$ (Table \ref{table_1}). Four spiral resonators at different frequencies are measured in reflection by inductively coupling them to a shared shorted transmission line as shown in Figure \ref{fig:exp_setup}(b). In addition to the four spiral resonators, our design features a co-planar waveguide (CPW) resonator with width = 20 $\mu$m and gap = 10 $\mu$m  that we use to benchmark our design and the fabrication process.

The devices are fabricated on a 50 nm Nb film deposited on an intrinsic silicon substrate (Star Cryoelectronics). Figures \ref{fig:exp_setup}(c-d) show SEM images of a spiral resonator with a 500 nm pitch (See supplementary materials for more details). 

\begin{table*}[t]
    \centering
    \caption{Designed and experimentally extracted parameters of the spiral resonators. We calculate $f_{g}$ and $L_{g}$ from Equations 1 and 2. The $Z_c  (= 2 \pi f_g L_g)$ calculation is described in the main text. $f_0$ and $\alpha$ are the fit results from the QP analysis with Equation 3. $Q_{\rm {TLS,0}}$ is the fit result from the TLS analysis with Equations 5 and 6. For $f_0$, $\alpha$, and $Q_{\rm {TLS,0}}$, the first value is from the fit to the internal quality factor $Q_{\rm {int}}$
, and the second value is from the fit to the resonance frequency $f_0$. }
    \begin{tabular}{c c c c c c c c c}
    \hline\hline
    label & p($\mu$m) & $Z_c(k\Omega)$ & $L_{g}$(nH) & $f_{g}$(GHz) & $f_{0}$(GHz) & $\alpha(\%)$ & $L_{k}$(nH) & $Q_{\rm {TLS,0}}(\times10^5)$ \\
    \hline
    R0 & 0.3 & 7.09 & 244 & 4.61 & 4.04 & 3.2-0.5 & 8.0-1.2 & 0.36 - 3.2 \\
    R1 & 0.3 & 6.67 & 205 & 5.19 & 4.53 & 1.5-0.1 & 3.2-0.3 & 1.2 - 7.4 \\
    R2 & 0.5 & 5.00 & 142 & 5.61 & 5.00 & none-8.0 & none-12 & none-2.7 \\
    R3 & 1.0 & 3.25 & 78 & 6.67 & 5.95 & 6.0-5.5 & 5.0-4.5 & 2.3 - 7.2 \\
    \hline\hline
    \end{tabular}   
   
    \label{table_1}
\end{table*}

\section{Measurements and Characterization}

The fabricated chips are mounted on a sample board and cooled down in a dilution refrigerator at milli-Kelvin temperatures. Figure \ref{fig:exp_setup}(e) shows a typical reflection measurement of a spiral resonator with a 2$\pi$ phase shift indicative of the overcoupled regime ($Q_{\rm {int}} > Q_{\rm {e}}$). From a fit to the complex $S_{11}$ signal we extract an internal quality factor $Q_{\rm {int}}$ and the fundamental resonance frequency. Next, we will characterize the spirals' geometric inductance and main loss mechanisms from reflection measurements at different powers and temperatures. We will then test their magnetic field resilience in a parallel magnetic field.

\subsection{Quantifying the Total Inductance}
 
The geometric inductance of the spiral resonators can be extracted by measuring how the fundamental resonance frequency $f_0=\frac{1}{2\pi \sqrt{LC}}$ shifts with temperature due to changes in the inductance. The total inductance is $L=L_g+L_k$ where $L_{g}$ and $L_{k}$ are the geometric inductance and the kinetic inductance of the superconducting spiral wires respectively. At higher temperatures, the environment has more energy to break up Cooper pairs, leading to a reduced Cooper pair density. As a result, the kinetic inductance increases with temperature while the geometry inductance remains constant.  The resulting temperature-dependent frequency shift can be modeled by Bardeen–Cooper–Schrieffer theory \cite{gao2008_equivalenceQPTempExtpairbreak,peruzzo2020_GemometricSI}:

\begin{figure}[h]
    \centering
    \includegraphics[width=\linewidth]{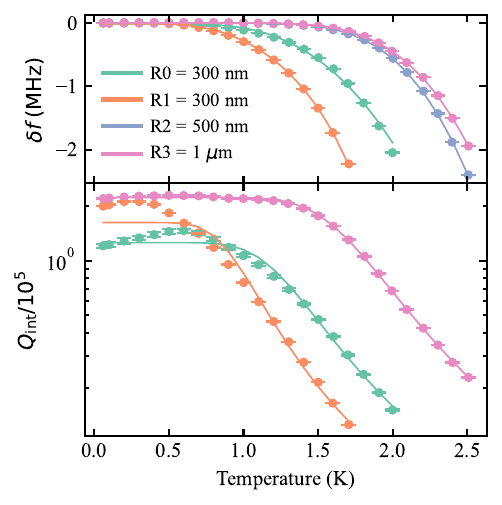}
    \caption{Shifts in the spiral frequency and quality factor characteristic of QPs. (Top) Frequency shift $\delta f$, and (bottom) internal qualify factor $Q_{\rm int}$ of the resonators as temperature is varied. Solid lines are the fits to Equation 3. Error bars of $\delta f$ and $Q_{\rm int}$ at each data point denote the statistical error, which is estimated by fitting a theoretical model for $ S_{11}$ as shown in Figure 1(e).
}
    \label{fig:temp_sweep}
\end{figure}
\begin{subequations}
\begin{align}
    \frac{\delta f(T)}{f_{0}(0)} &= -\frac{\alpha\gamma}{2}\frac{\delta\sigma_{2}(T,\Delta )}{\sigma_{2}(T,\Delta )} \label{eq:QP freq}\\
\intertext{and}
    \delta Q^{-1}_{\text{int}}(T) &= \alpha\gamma\frac{\delta\sigma_{1}(T,\Delta )}{\sigma_{2}(T,\Delta )},
    \label{eq:QP Qint}
\end{align}
\end{subequations}

where $\delta f(T) = f_{0}(T)-f_{0}(0) $ is the relative fundamental resonance frequency shift; $\delta Q^{-1}_{\rm int}= Q^{-1}_{\rm int}(T)-Q^{-1}_{\rm int}(0)$ is the relative internal quality factor shift;
$\alpha=L_k / \left( L_k + L_g \right)$ is the fraction of the kinetic inductance $L_{k}$ to total inductance $L_{k}+L_{g}$; $\gamma$ is a material and film dependent parameter (for thin film Nb, $\gamma$ = -1) \cite{gao2008_physicsMicrowaveResonators};  $\sigma_{1}$ and $\sigma_{2}$ are the real and imaginary parts of the total conductivity $\sigma$ of the superconducting film respectively; and $\Delta = 1.76k_{B}T_{c}$  \cite{gao2008_physicsMicrowaveResonators} is the superconducting gap of Nb where $k_{B}$ and $T_{c}$ are the Boltzmann constant and the critical temperature of the Nb film respectively. 

From $L_g$ we calculate  the characteristic impedance $Z_c = 2 \pi f_g L_g$ and find $Z_c = 3.25$ k$\Omega$ for the spiral resonator with the largest pitch and $Z_c = 7.09 $ k$\Omega$ for the spiral resonator with the smallest pitch (Table \ref{table_1}). We extract the ratio of kinetic to total inductance, $\alpha$, by fitting the $f_0$ data to Equation \ref{eq:QP freq} with $f_{0}(0),\alpha$ and $T_{c}$ as the fitting parameters and by fitting the $Q_{\rm int}$ data to Equation \ref{eq:QP Qint} with $Q_{\rm int}(0),\alpha$ and $T_{c}$ as the fitting parameters (Table 1). We find $\alpha$ values ranging from 0.1-8$\%$ as expected for resonators with inductances primarily due to geometric inductance.
 R2 shows reflection measurements that could not be modeled with a $S_{11}$ Lorentzian dip (Supplementary Figure 1). We hypothesize that this is due to impedance mismatches that cannot be corrected in reflection measurements \cite{mcrae_2020_materialsLossMeasurementsMicrowaveResonators}. Therefore, the R2 $Q_{\rm int}$ values are not included in Figure \ref{fig:temp_sweep}. 

Now we compare the extracted $\alpha$ and $T_c$ values obtained with Equation 3a to those obtained with Equation 3b [Supplementary Figure 3(a)]. We find that the $f_0$ fit results show $\alpha$ values of 0.5$\%$, 0.1$\%$, 8.0$\%$, and 5.5$\%$ for R0, R1, R2 and R3 respectively. We independently extract $\alpha$ from $Q_{\rm int}$ fits and find $\alpha$'s between 1.5-6$\%$ (Table \ref{table_1}). We attribute the discrepancy between the two methods to the sensitivity of $Q_{\rm int}$ to two-level systems. At lower temperatures, the changes in $f_0$ and $Q_{\rm int}$ with temperature are due to the effects of TLSs,  which are not accounted for in Equation 3. If this low-temperature change is small compared with the overall change in $f_0$ and $Q_{\rm int}$ caused by the high-temperature quasiparticles (QPs), the TLS effect on $f_0$ and $Q_{\rm int}$ is masked by the QP losses, and so the fits are robust to the low-temperature changes. We hypothesize that for R0 and R1, the fits are not as robust to TLS effects as they are for R3, thus giving different results for $\alpha$ and $T_c$ for the $f_0$ and $Q_{\rm int}$ fits. Despite these uncertainties, we can conclude that for our resonators the majority of the inductance is due to the geometric contribution.

Next we compare the extracted $T_c$ to four-point measurements [Supplementary Figure 3(a)]. We find that the $f_0$ fit results show $T_c$ values of 3.5 K, 2.1 K, 7.9 K, and 7.9 K for R0, R1, R2 and R3 respectively. For $Q_{\rm int}$ fits, these values are 5.4 K, 3.9 K, and 8.3 K (for R0, R1 and R3 respectively). We attribute the deviations in $T_c$ between the two fits to the above mentioned effects of low-temperature TLSs. Four-point measurement $T_c$ values are 7.5 K, 7.9 K, and 8.1 K for 50 $\mu$m long wire with wire width of 150 nm (designed wire width for R0 and R1), 500 nm (designed wire width for R2), and 1 $\mu$m (designed wire width for R3) respectively. We attribute the differences in $T_c$ to imperfections in spiral fabrication, making the spiral wire dimensions different from the designed wire widths. Additionally, for small features such as our spirals, the value of $T_c$ is sensitive to geometry, making the $T_c$ values of four-point measurements different from the spiral $T_c$ values. 
To extract the characteristic impedance from the measurements, we estimate the kinetic inductance $L_{k}$ for each resonator by combining the fitted $\alpha$ with our analytical estimate for the geometric inductance based on the designed dimensions (Equation \ref{eq:Lg}).  We find kinetic inductances on the order of a few nH and calculate geometric inductances between 78 and 244 nH (Table \ref{table_1}). 

We do a self-consistency check for the extracted spiral parameters by comparing the resonance frequency $f_0$ to the designed analytical frequency $f_g$. The analytical model, which only considers the geometric inductance, can be adjusted to include the kinetic inductance:  $f_0 =f_g\sqrt{L_g / (L_k + L_g)}$ \cite{peruzzo2020_GemometricSI}. For each spiral, we calculate $f_g$ by substituting $L_k$,  $L_g$, and the fitted $f_0$   in this equation. We find these estimates of $f_g$ to be 7 - 13$\%$ lower than the $f_g$ calculated from the designed dimensions (Equation \ref{eq:fg}).  One possible explanation for the discrepancy is variations in spiral dimensions due to an imperfect fabrication processes. When the fabricated spirals do not have the designed gap ($s$) to wire ($w$) ratio of 1:1, the current-sheet method (Equation \ref{eq:Lg}) gives a maximum of 8$\%$ error in $L_g$ for $s\leq3w$ \cite{Mohal_1999_planar_spital_inductances}. We confirm that the fabricated spirals satisfy this geometric condition with SEM images. Other possible sources for the discrepancy might be the change in self-capacitance due to deviations in the $s:w$ ratio of the fabricated spirals or parasitic capacitance and inductance to the transmission line and ground plane.  

\subsection{Characterization of Loss Mechanisms }

\begin{figure}[h]
    \centering
    \includegraphics[width=\linewidth]{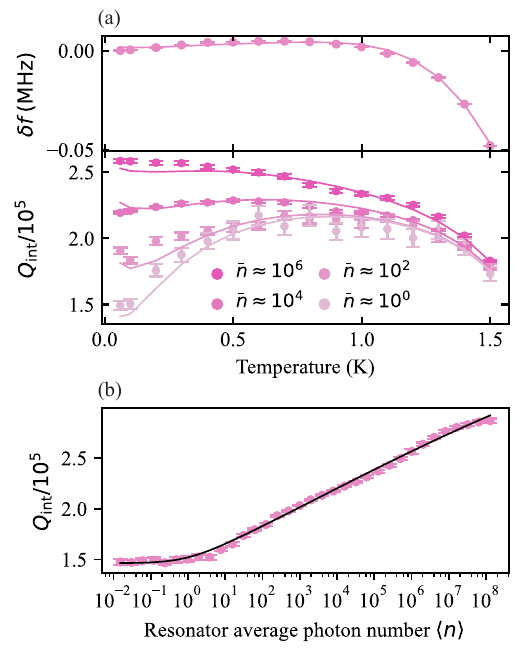}
    \caption{Shifts in the spiral frequency and quality factor characteristic of TLS.
 (a) (Top) Frequency shift $\delta f$, (bottom) internal qualify factor $Q_{\rm int}$ of the resonator (R3 in the text) as temperature is varied. Solid lines represent fits to Equation 5. (b) $Q_{\rm int}$ as resonator average photon number is varied. The solid line represents the fit to Equation \ref{eq:TLS_ncrit_loss}. }
    \label{fig:main}
\end{figure}

Having determined the impedance of the fabricated spiral resonators, next we characterize their dissipation mechanisms. We distinguish three common sources of loss for superconducting resonators arising from TLSs, QPs, and other temperature and power-independent losses for example radiative decay or packaging.
First, we measure the power-dependent quality factor which shows behavior characteristic of TLS loss: the quality factor saturates to a single photon level at low powers and the TLS bath becomes saturated by the photons in the resonator as evidenced by a rise in $Q_{\text{int}}$ in the high power regime [Figure \ref{fig:main}(b)]. We fit the data into a simplified model for TLS-induced dielectric loss \cite{gao2008_physicsMicrowaveResonators,burnett_2017_NW_neonFIB}:
\begin{equation}
    Q_{\text{int}}^{-1}= Q_{\text{TLS,0}}^{-1}\frac{\tanh{\left(\frac{hf_0}{2k_BT}\right)}}{(1+\frac{\langle n \rangle}{n_c})^{\beta}}+Q_{\text{other}}^{-1},
    \label{eq:TLS_ncrit_loss} 
\end{equation}

where $Q_{\rm {TLS,0}}$ is a temperature-independent TLS quality factor, $\langle n \rangle$ is the average number of photons in the resonator,  $n_c$ is the critical number of photons in the resonator, the exponent $\beta$ is a parameter to account for TLS interactions and $Q_{\text{other}}$ takes into account all non-TLS related loss. A typical power dependence curve is shown in Figure \ref{fig:main}(b) and we find single photon quality factors $Q_{\text{int}}$ between $0.9\text{x} 10^5$ and $2.1\text{x} 10^5$ for our spiral devices (Supplementary Figure ).

Next, we study the temperature- and power-dependent noise, which requires a more comprehensive noise model to account for the temperature dependence of the TLS loss tangent and QPs. Following \cite{deLeon_2023_Disentangling_Ta}, we will model the resonance frequency shift and the  internal quality factor shift:

\begin{subequations}
\begin{align}  
    \frac{\delta f(T)}{f_{0}(0)} &= {\left({\frac{\delta f(T)}{f_{0}(0)}}\right)}_{\text{TLS}} {+}  {\left({\frac{\delta f(T)}{f_{0}(0)}}\right)}_{\text{QP}}\label{eq:TLS_QP_freq}\\ 
\intertext{and}
    Q_{\text{int}}^{-1}&= Q_{\text{TLS}}(\langle n \rangle,T)^{-1}+Q_{\text{QP}}(T)^{-1}+Q_{\text{other}}^{-1}.
    \label{eq:TLS_QP_Qint}
\end{align}
\end{subequations}

Here the TLS resonance frequency shift and the quality factor are \cite{pappas_2011_TLSlosses,deLeon_2023_Disentangling_Ta}
\begin{subequations}
\begin{align} 
&\begin{aligned}
    &{\left({\frac{\delta f(T)}{f_{0}(0)}}\right)}_{\text{TLS}}=\frac{1}{\pi Q_{\text{TLS,0}}}{\text{Re}}\left[ \Psi\left( \frac{1}{2}{+}i \frac{h f_{0}}{ 2\pi k_B T}
    \right) \right. \\
    &\qquad\qquad\qquad - {\text{ln }\left( \frac {hf_{0}}{2\pi k_B T}  \right)}     \left.\vphantom{\frac{1}{2}} \right]
\end{aligned}
\label{eq:TLS_freq}\\
\intertext{and}
&Q_{\text{TLS}}(\langle n \rangle,T)= Q_{\text{TLS,0}} \frac{\sqrt{1+(\frac{\langle n \rangle^{\beta_2}}{DT^{\beta_1}})\tanh{\left(\frac{hf_0}{2k_BT}\right)}}}{\tanh{\left(\frac{hf_0}{2k_BT}\right)}}.
\label{eq:TLS_Qint}
\end{align}
\end{subequations}
Now the TLS quality factor has temperature and photon number dependence. Here $\Psi$ is the complex digamma function; and $D$, $\beta_1$ and $\beta_2$ are parameters that characterize the TLS saturation.
${\left({\frac{\delta f(T)}{f_{0}(0)}}\right)_{\rm QP}}$ and $Q_{\rm QP}$ are the resonance frequency shift and the quality factor associated with QP losses given in Equation 3.

 We extract $Q_{\rm {TLS,0}}$ by fitting the $Q_{\rm int}$ data to Equation \ref{eq:TLS_QP_Qint} with $Q_{\rm {TLS,0}}$,  $\alpha$, $T_c$, $D$, $\beta_{1}$, $\beta_{2}$ and $Q_{\rm other}$ as fitting parameters; and by fitting the $f_0$ data with Equation \ref{eq:TLS_QP_freq} with $Q_{\rm {TLS,0}}$,  $\alpha$, $T_c$, and $f_0(0)$ as fitting parameters. At low temperatures, the internal quality factor is limited by the TLSs for low photon number, while at high enough excitation power, the TLS bath is saturated and the quality factor is dominated by residual losses $(Q_{\rm other}$).  At higher temperatures, the losses are dominated by QPs independent of power. Therefore we hypothesize that at low temperatures the temperature and power sweeps, modeled with Equation 5, do not provide accurate $\alpha$ and $T_c$.

We compare $Q_{\rm {TLS,0}}$ of the $f_0$ and $Q_{\rm int}$ fits [Supplementary Figure 3(b)]. We find that $Q_{\rm{int}}$ fit results are significantly lower than $f_0$ fits for R0, R1, and R3 respectively. We hypothesize that masking of the TLS-induced $f_0$ shift by the QP-induced $f_0$ shift, along with having fewer data points than $Q_{\rm int}$ fits due to the power independence of the $f_{0}$ shift, cause these deviations between the two fits.

\subsection{Magnetic Field Compatibility}

Next, we measure the spiral resonators' characteristics in the presence of an applied magnetic field.
\begin{figure}
    \centering
    \includegraphics[width=\linewidth]{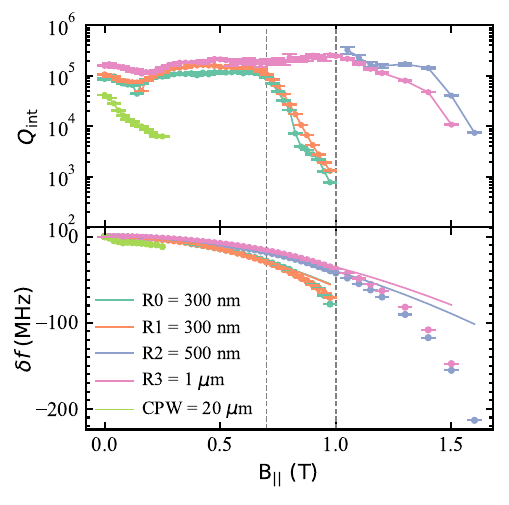}
    \caption{Performance of the spiral resonators in the external magnetic field $B_{\Vert}$. For comparison CPW resonator performances are included. (Top) The internal quality factor $Q_{\rm int}$ plotted against $B_{\Vert}$. The solid lines are guides to the eye.
    (Bottom) The resonance frequency $f_0$ plotted against $B_{\Vert}$. Quadratic fits for each spiral resonator are shown with solid lines. Vertical dashed lines indicate where $Q_{\rm int}$ starts to degrade drastically due to vortex effects.}

    \label{fig:MagneticFieldS_Sweep}
\end{figure}

\begin{figure}[h]
    \centering
    \includegraphics[width=\linewidth]{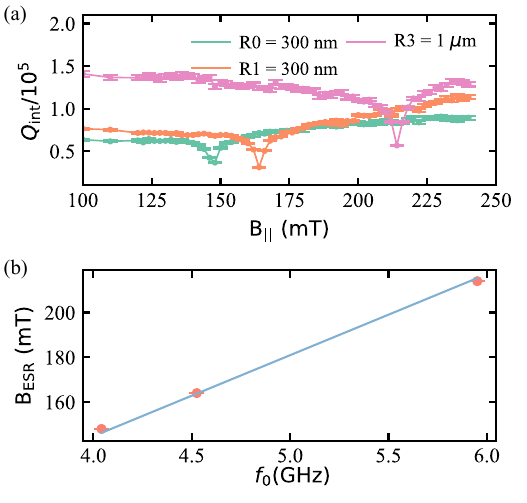}
    \caption[Performance of spiral resonators in external magnetic fields $B_{\Vert}$ near the ESR dip.]
    {Performance of spiral resonators in external magnetic fields $B_{\Vert}$ near the ESR dip. 
    (a) The internal quality factor $Q_{\rm int}$ of the resonators near the electron spin resonance condition. (b) $B_{\Vert}$ of each resonator where the $Q_{\rm int}$ dip occurs, plotted against its respective resonance frequency $f_{0}$. The solid blue line is the fit to $h f_0 = g \mu_{B} B_{\Vert}$ as described in the text.}
    \label{fig:ESR_Sweep}
\end{figure}

For all the spirals, the extracted internal quality factor $Q_{\rm int}$ is  $\sim 10^{5}$ at the zero magnetic field. The spiral with larger pitch (R3) retains this higher $Q_{\rm int}$ up to higher in-plane field than those with smaller pitch (R0 and R1). We benchmark the performance of the spirals in the external magnetic field by comparing them to simultaneous measurements of the CPW resonator. We see that the spirals have better magnetic field performance than the CPW resonator. We attribute this to the suppression of vortices with decreasing device width.  We note, however, that despite having a larger wire width than R0 and R1, R3 has better magnetic field resilience. We attribute this to the fact that smaller wire widths are more sensitive to impurities, leading to a decrease in critical magnetic field strength independent of vortex suppression \cite{kwon2018_MagneticNbMicrostrip}. Finally, as discussed above, the low-magnetic field reflection measurements for R2 do not show a Lorentzian dip, which we attribute to impedance mismatches. We do, however, observe a Lorentzian dip for magnetic fields larger than $\sim$1 T, permitting the extraction of $Q_{\rm {int}}$ [Figure \ref{fig:MagneticFieldS_Sweep} (top)]. We interpret this as the impedance mismatch being reduced in the presence of a strong magnetic field.

One feature apparent in Figure \ref{fig:MagneticFieldS_Sweep} is a sharp dip in $Q_{\rm int}$ around 100 - 200 mT. We investigated this feature by first warming the sample up above Nb transition temperature and cooling it back down, thus eliminating any magnetic field hysteresis effects caused by the previous magnetic field sweep. We then measured a higher-resolution magnetic field sweep in that region and show the results in Figure \ref{fig:ESR_Sweep}. We attribute this dip to the electron spin resonance (ESR) caused by the resonator coupling to the magnetic impurities in the substrate and the other interfaces \cite{Vandersypen2016_KineticInductanceNWs}. In the presence of a static magnetic field, these impurities will have a Zeeman splitting with an energy difference of $g\mu_{B}B_{\Vert}$, where $g$ and $\mu_{B}$  are the Land$\rm\acute{e}$ g-factor and the Bohr magneton respectively. As shown in Figure \ref{fig:ESR_Sweep}(a) ,  at the ESR condition of $h f_0=g\mu_{B}B_{\Vert}$ where  $h$ is the Planck constant, there is a sharp increase in the resonator losses indicated by the $Q_{\rm int}$ minima. In Figure \ref{fig:ESR_Sweep}(b) we plot the relationship between the magnetic field strengths $B_{\Vert}$ at which these minima occur and the corresponding resonance frequency $f_{0}$ of each resonator.

We fit this data to the Zeeman resonance condition $h f_0=g\mu_{B}B_{\Vert}$ and obtain a Land$\rm \acute{e}$ g-factor of 1.97$\pm$0.01, consistent with the hypothesis that the dips are due to the resonators coupling to free electron spins. The spirals show an enhancement in $Q_{\rm int}$ with an increasing magnetic field following the ESR dip. There are two hypotheses regarding this phenomenon: the Abrikosov vortices trap the lossy QPs or there is a contribution from magnetic impurities to the resonator losses before the ESR dip  \cite{zollitsch2019_tuningQResMagfield}.

The bottom graph of Figure \ref{fig:MagneticFieldS_Sweep} shows the shift in $f_{0}$ against $B_{\Vert}$. At lower magnetic fields, the decrease in $f_{0}$ with increasing $B_{\Vert}$ is due to the increase in  $L_{k}$ as a consequence of a decreasing Cooper pair density. This can be fitted to a quadratic function as shown by the solid lines in 
  Figure \ref{fig:MagneticFieldS_Sweep} \cite{Vandersypen2016_KineticInductanceNWs,Kouwenhoven2019_BResilientCPW}. At higher magnetic fields, the frequency shift deviates from the quadratic trend when the resonator's dominant loss mechanism shifts from QPs to vortex losses \cite{zollitsch2019_tuningQResMagfield,kwon2018_MagneticNbMicrostrip,Martinis2009_VorticesSC}. The magnetic field response of the vortices is characterized by a complex resistivity, with the real part contributing to vortex-related losses and the imaginary part corresponding to their reactive response \cite{Martinis2009_VorticesSC}. When the frequency shift is dominated by vortices but not by the QPs generated by the external magnetic field, the frequency shift deviates from the quadratic form. Here, we have fitted the data to the quadratic model until each resonator's quality factor starts to decrease drastically, as indicated by vertical grey lines in Figure \ref{fig:MagneticFieldS_Sweep}. We see that the frequency shift deviates from the quadratic function once the spirals start to decay, confirming that the losses arise from the vortices. In contrast to the spirals, we see that $Q_{\rm int}$ of the CPW resonator decreases drastically with increasing magnetic field from the beginning. This indicates that the dominant losses arise from the vortices from the start. We confirm this further by observing that the decrease in $f_{0}$ cannot be fitted to a quadratic function in a similar manner to spirals [Figure 4(bottom)].

\section{Conclusions}
In this work we have fabricated high-impedance Nb geometric resonators with impedances of 3-7 k$\Omega$. In the single photon limit these resonators show a quality factor of $\sim 10^5$, which they retain in external in-plane magnetic fields up to $\sim$ 1 T. Further in the study, we probe resonator loss mechanisms by varying the temperature and photon number and measuring the resonator response. We confirm that at low temperatures, the losses are dominated by TLS losses. At high temperatures and powers the TLSs saturate, and therefore no longer contribute to the resonator losses. In this regime we identify QPs as the dominant loss mechanism. 

Our results demonstrate the feasibility of operating high-impedance superconducting resonators in strong magnetic fields. Because our fabrication methods are compatible with existing material platforms and architectures, we expect that these resonators could enable hybrid quantum devices. In particular we propose leveraging the high impedance and high magnetic field resilience realized in our spiral resonators to coherently couple microwave photons to electron spin qubits. Furthermore, high-impedance Nb resonators can be easily integrated into more conventional circuit QED systems to explore novel qubit architectures intrinsic noise protection and high coherence.

\section{Acknowledgment}
This work is supported by the National Science Foundation under Grant No. 2210309.

%
\newpage
\onecolumngrid
\section*{Supplementary Information}
\setcounter{section}{0}
\section{Chip Fabrication}

A 50 nm Nb film was deposited on a 4" double-sided polished intrinsic silicon wafer by Star Cryogenics. Then, the wafer was diced into 10 mm$\times$ 10 mm square chips. Then a diced chip was covered with $\sim $ 450 nm thick MMA EL 11 spun at 4200 rpm. After spinning the resist, the chip was baked at $140-150^\circ$ C for 10 minutes. The spirals along with the CPW resonator and the on-chip 2D transmission lines were patterned with Elionix ELS-S50EX. After the exposure, the chip was developed in MIBK/IPA 1:3 developer at $2^\circ$ C for 1 minute and then rinsed with IPA to stop the development. Next, the reactive iron etch was done with South Bay Technology Reactive Ion Etcher RIE-2000. The etching was done with $\rm {Ar/SF_{6}}$ chemistry for 1-2 minutes at 80 W forward power. The resist was removed with hot Remover PG, followed by a sonication and an IPA wash. Then, the chip was mounted to a copper box with GE Varnish, wire bonded to a printed circuit board and cooled down.

\section{Reflection measurements of R2}
The supplementary Figure 1 shows the reflection measurement of R2 at 60 mK and at 2.5 K at zero external magnetic field. We see that at 60 mK the shape of the reflection measurement is not a dip but with increasing temperature, the shape transforms into a Lorentzian dip which we can successfully fit to get $Q_{\rm int}$. Similar behavior is observed with increasing magnetic field at 60 mK. 

 \section{spiral characteristic shift due to the effect of TLS}

The main text shows R3 resonator characteristic shift with temperature and power due to TLS. The supplementary Figure 2 shows R0, R1 and R2 characteristic shifts due to TLS which are not shown in the main text. 

\section{Comparison of $Q_{\rm int}$ and $f_0$ fit results}
 
 The supplementary Figure 3(a) shows the extracted $\alpha$ and $T_c$ results of the fits shown in the main text Figure 2. Four-point measurement results of Nb wires with wire widths as same as the spiral wire widths are also included in Figure 3(a). The R2 results are excluded as it is giving a non-positive number for $Q_{\rm int}$. The supplementary Figure 3(b) shows the extracted $Q_{\rm TLS,0}$ results of the fits shown in the main text Figure 3 and the supplementary Figure 2.
 \begin{figure}[h]
    \centering
    \includegraphics [width=\linewidth] {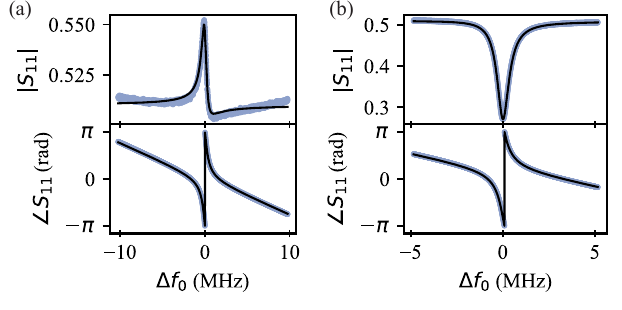}
    \caption{Magnitude (top) and phase (bottom) of R2 at (a) 60 mK and (b) 2.5 K. Measured data are shown in pastel blue. The magnitude of the measurement has a Lorentzian-shaped dip at 2.5 K. A theoretical model for $S_{11}$ is shown in black.}
    \label{fig:R2}
\end{figure}
\clearpage
\begin{figure*}[p]
    \centering  \includegraphics{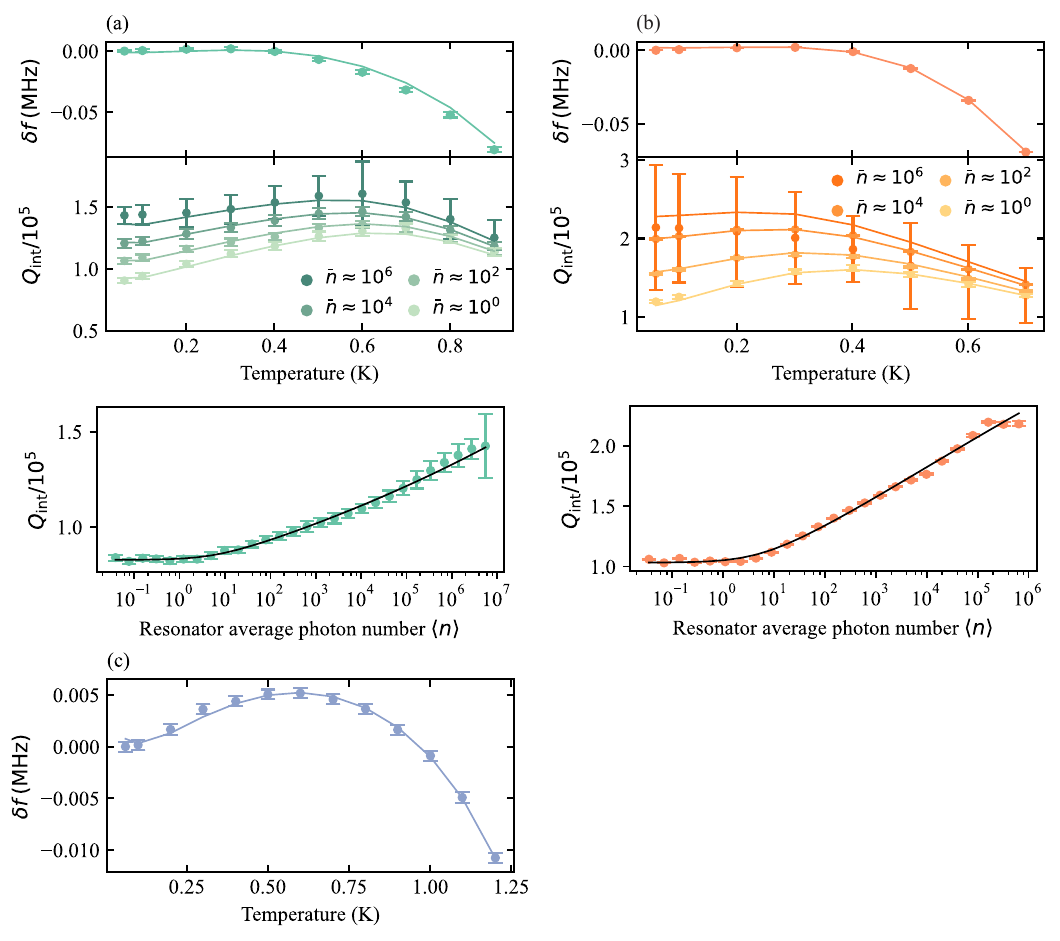}   
    \caption{Frequency shift $\delta f$ and internal qualify factor $Q_{\rm int}$ (top) as temperature is varied and (bottom) average number of photons in the resonator is varied of (a) R0, (b) R1, and (c) R2. Solid lines represent fits to (top) Equation 5 and (bottom) Equation 4 in the main text.}
    \label{fig:Qi vs F}
\end{figure*}

\begin{figure*}[p]
    \centering
    \includegraphics{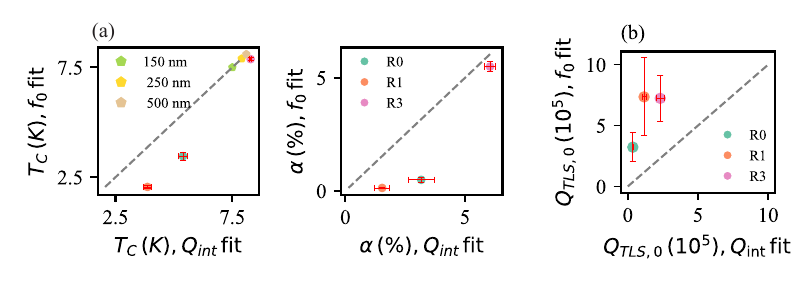}
    \caption{The extracted parameters from the spiral resonator fits. (a) $\alpha$ and $T_c$ are obtained with Equation 3 in the main text. The $T_c$ values obtained with four-point measurements are also included in the graph (pentagon-shaped markers). (b) $Q_{\rm TLS,0}$ obtained with Equations 5 and 6 in the main text.}
    \label{fig:Qi vs F}
\end{figure*}

\end{document}